\pdfoutput=1

\documentclass[iop, apj, twocolappendix, numberedappendix]{emulateapj}

\usepackage{txfonts}
\usepackage{natbib}
\usepackage[colorlinks=true,urlcolor=blue,linkcolor=blue,citecolor=blue]{hyperref}

\newcommand{\mpch}{\>h^{-1}{\rm {Mpc}}}

\newcommand{\msunh}{\>h^{-1} M_\odot}

\def\LCDM{$\Lambda$CDM\ }

\def\mvir{M_{\rm vir}}

\def\gcm3{\mathrm{g} / \mathrm{cm}^3}
\def\m200m{M_{\rm 200m}}

\def\gtsima{$\; \buildrel > \over \sim \;$}
\def\ltsima{$\; \buildrel < \over \sim \;$}
\def\prosima{$\; \buildrel \propto \over \sim \;$}
\def\gsim{\lower.7ex\hbox{\gtsima}}
\def\lsim{\lower.7ex\hbox{\ltsima}}
\def\simgt{\lower.7ex\hbox{\gtsima}}
\def\simlt{\lower.7ex\hbox{\ltsima}}
\def\simpr{\lower.7ex\hbox{\prosima}}

\def\rhom{\rho_{\rm m}}

\usepackage{etoolbox}
\makeatletter
\patchcmd{\NAT@citex}
  {\@citea\NAT@hyper@{\NAT@nmfmt{\NAT@nm}\NAT@date}}
  {\@citea\NAT@nmfmt{\NAT@nm}\NAT@hyper@{\NAT@date}}
  {}
  {}
\patchcmd{\NAT@citex}
  {\@citea\NAT@hyper@{%
     \NAT@nmfmt{\NAT@nm}%
     \hyper@natlinkbreak{\NAT@aysep\NAT@spacechar}{\@citeb\@extra@b@citeb}%
     \NAT@date}}
  {\@citea\NAT@nmfmt{\NAT@nm}%
   \NAT@aysep\NAT@spacechar%
   \NAT@hyper@{\NAT@date}}
  {}
  {}
\patchcmd{\NAT@citex}
  {\@citea\NAT@hyper@{%
     \NAT@nmfmt{\NAT@nm}%
     \hyper@natlinkbreak{\NAT@spacechar\NAT@@open\if*#1*\else#1\NAT@spacechar\fi}%
       {\@citeb\@extra@b@citeb}%
     \NAT@date}}
  {\@citea\NAT@nmfmt{\NAT@nm}%
   \NAT@spacechar\NAT@@open\if*#1*\else#1\NAT@spacechar\fi%
   \NAT@hyper@{\NAT@date}}
  {}
  {}
\makeatother

\shorttitle{The Fabric of the Universe}
\shortauthors{Diemer and Facio}
\journalinfo{Publications of the Astronomical Society of the Pacific, 129:058013 (10pp), 2017 May}
\submitted{Received 2017 March 2; accepted 2017 March 29; published 2017 April 17}

\begin{document}


\def\figdir{.}
\def\widththree{5.9cm}


\title{The fabric of the universe:\\ exploring the cosmic web in 3D prints and woven textiles}
\author{Benedikt Diemer\altaffilmark{1,2} and Isaac Facio\altaffilmark{3,4}}

\affil{
$^1$ Institute for Theory and Computation, Harvard-Smithsonian Center for Astrophysics, 60 Garden St., Cambridge, MA 02138, USA; {\tt benedikt.diemer@cfa.harvard.edu}\\
$^2$ Department of Astronomy and Astrophysics, The University of Chicago, Chicago, IL 60637, USA \\
$^3$ Department of Fiber and Material Studies, School of the Art Institute of Chicago, 36 S Wabash Ave., Chicago, IL 60603, USA \\
$^4$ Department of Textiles, The Art Institute of Chicago, 111 S Michigan Ave., Chicago IL 60603, USA \\
}


\begin{abstract}
We introduce {\it The Fabric of the Universe}, an art and science collaboration focused on exploring the cosmic web of dark matter with unconventional techniques and materials. We discuss two of our projects in detail. First, we describe a pipeline for translating three-dimensional density structures from $N$-body simulations into solid surfaces suitable for 3D printing, and present prints of a cosmological volume and of the infall region around a massive cluster halo. In these models, we discover wall-like features that are invisible in two-dimensional projections. Going beyond the sheer visualization of simulation data, we undertake an exploration of the cosmic web as a three-dimensional woven textile. To this end, we develop experimental 3D weaving techniques to create sphere-like and filamentary shapes and radically simplify a region of the cosmic web into a set of filaments and halos. We translate the resulting tree structure into a series of commands that can be executed by a digital weaving machine, and present a large-scale textile installation.
\end{abstract}


\section{Introduction}
\label{sec:intro}

\begin{figure*}
\centering
\includegraphics[trim = 0mm 0mm 0mm 0mm, clip, scale=0.35]{\figdir/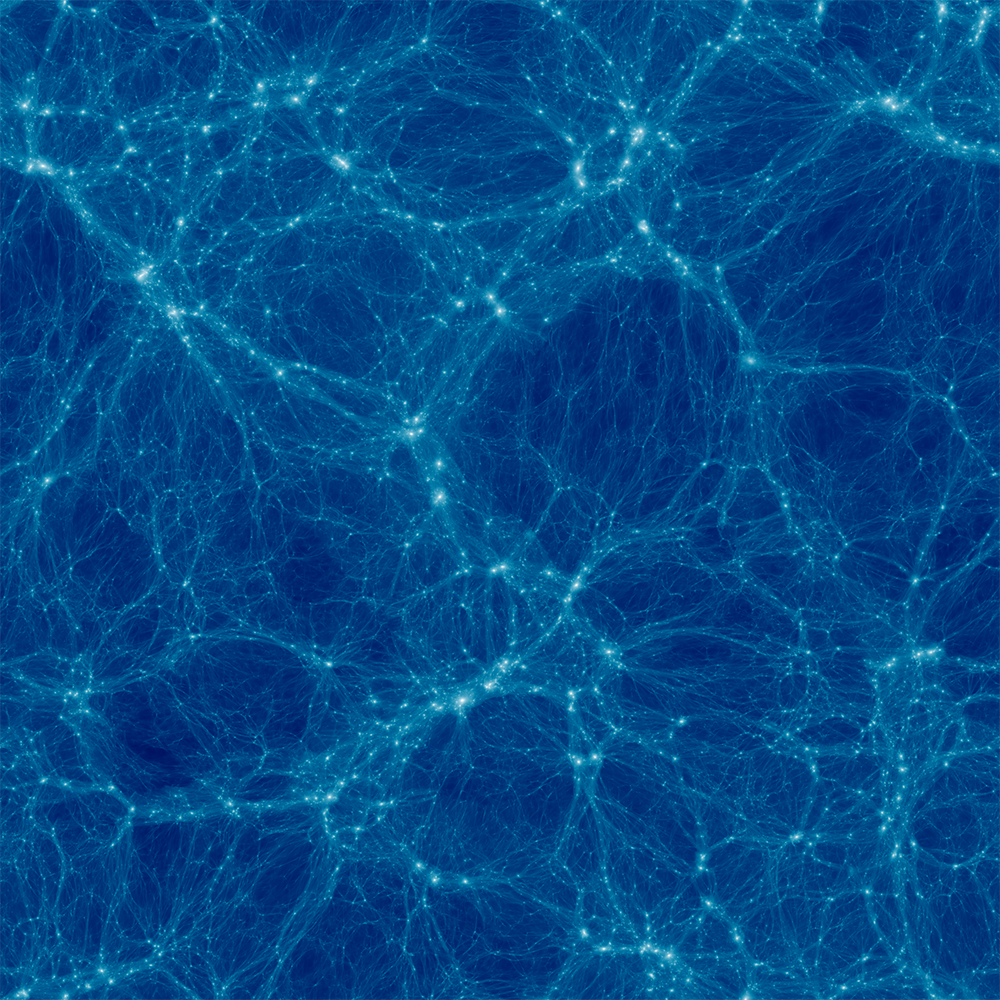}
\includegraphics[trim = 0mm 0mm 0mm 0mm, clip, scale=0.35]{\figdir/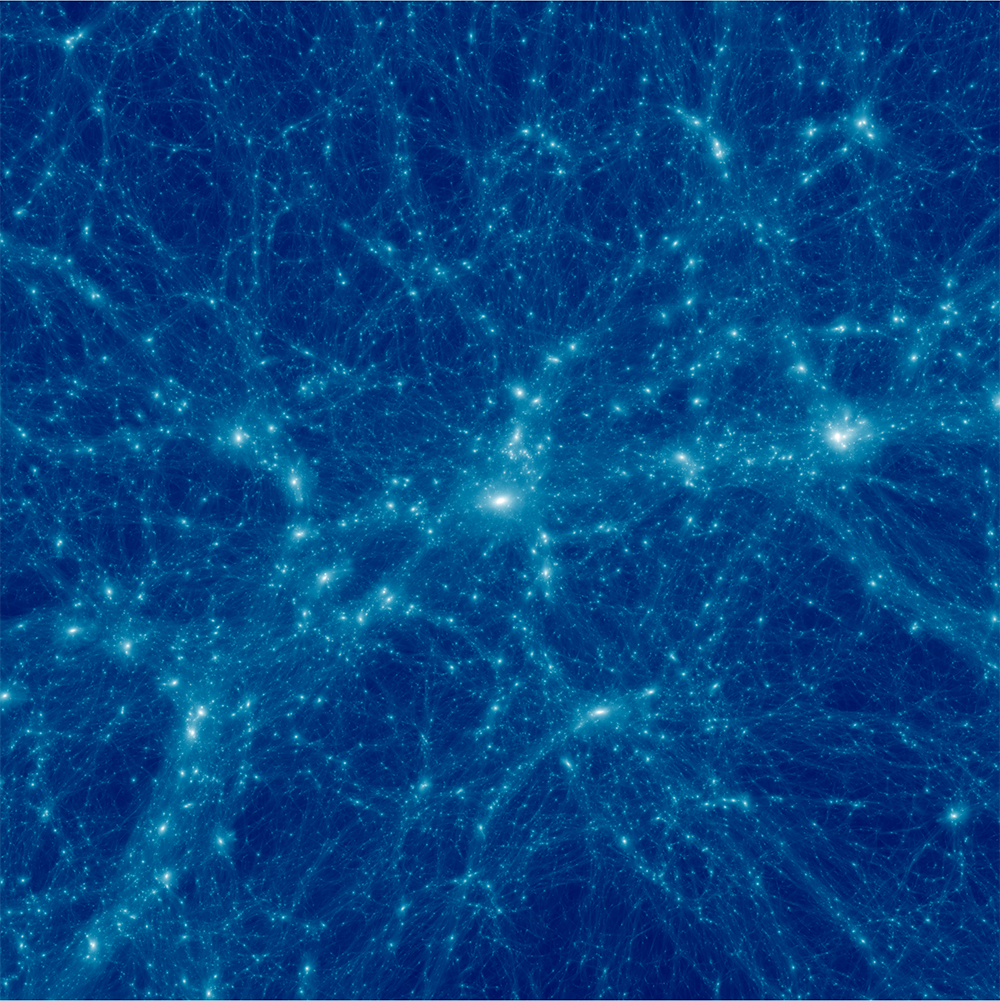}
\caption{Projected dark matter density in slices through the L0125 simulation. The left panel shows a slice through the entire simulation volume, $125 \mpch$ on a side, which will be referred to as the cosmological volume. The right panel shows the projection of a $50 \mpch$ cube around a massive halo, referred to as the infall region. The visualizations were created using the \textsc{GoTetra} code (P. Mansfield et al., in preparation). The thread-like appearance of filaments in this type of visualization motivated us to express the cosmic web as a textile.}
\label{fig:viz}
\end{figure*}

One of the predictions of the predominantly accepted $\Lambda$CDM paradigm of structure formation is that dark matter collapses into a cosmic web of walls, filaments, and halos \citep{zeldovich_70, peebles_80_structurebook}. Since the early days of research into the large-scale structure of the universe, numerous images of the visually striking appearance of the cosmic web have been created \citep[e.g.,][]{doroshkevich_78, springel_05_millennium, abel_12}. However, it immediately became clear that two-dimensional projections of the web cannot do its true, three-dimensional structure justice, leading to attempts at characterizing its elements in three-dimensional space. These investigations began with the iso-density surface of \citet[][Figure 4]{klypin_83}, endearingly nicknamed the ``chicken universe'' due to its resemblance to a certain bird species. Even with more advanced techniques for pseudo-3D visualizations \citep[e.g.,][]{bond_96_filaments}, the projection onto an image inevitably leads to a loss of information.

With the advent of 3D printing and virtual reality \citep{ferrand_16, vohl_16}, we finally have a chance to visualize cosmic web data in their true, three-dimensional nature. While limited in size and resolution, 3D printers have recently become widely accessible and the first 3D prints of astrophysical data have already been created, including subjects such as stellar winds \citep{madura_15}, the cosmic microwave background \citep{clements_17}, and general data visualization \citep{vogt_14, vogt_16}. Miguel Aragon-Calvo\footnote{See his website at \hyperlink{http://www.dataviz.science/}{www.dataviz.science}} printed a 3D model of a piece of the cosmic web in which one of the three axes represents time. In this paper, we focus on real-space structures and create 3D models of a large cosmological region and a zoomed region around a massive halo. Interestingly, we find that certain types of cosmic web structures (such as walls) are lost in projection but readily apparent in 3D prints.

However, the goal of the {\it Fabric of the Universe} collaboration is to go beyond the visualization of astrophysical data and to create art, meaning that we do not strive to find a medium that allows for the most faithful or accurate representation of dark matter simulations. Instead, we aim to reduce the cosmic web to its fundamental shapes and components, and to create installations that express those characteristics. Dark matter is a particularly fascinating subject for such an exploration because, unlike other astrophysical objects such as stars or galaxies, it cannot be seen. Artistic representation has the ability to provide shape to such abstract concepts so that they can be grasped immanently.

There are some thought-provoking precedents for artworks inspired by astronomical data. For example, Josiah McElheny and David Weinberg's sculpture {\it An end to modernity} arranges glass galaxies on the ends of metal spikes which are thrust out from a central hub, such that the distance of the glass elements from the center signifies the age of the universe since the Big Bang \citep[][see also David Weinberg's website\footnote{The website \hyperlink{astronomy.ohio-state.edu/~dhw/mcelheny}{astronomy.ohio-state.edu/\textasciitilde dhw/mcelheny} gives an excellent overview of their artistic process as well as follow-up sculptures that refined the initial concept.}]{weinberg_10}. The cosmic web itself has served as inspiration for a number of artists, for example Tom\'{a}s Saraceno who visually connects the structure of spider webs to the cosmic web in large-scale installations.\footnote{See Saraceno's website at \hyperlink{http://tomassaraceno.com/}{tomassaraceno.com}, e.g. his 2009 installation ``Galaxies forming along filaments, like droplets along the strands of a spider’s web.'' See also Nathan Kandus' work {\it Large-scale structure}, \hyperlink{nathankandus.com}{nathankandus.com}.} Compared to Saraceno's work, we strive to maintain a more stringent connection between the underlying simulation data and the final structure, although we allow for artistic freedom in how this representation is achieved. Our approach is more inspired by that of McElheny and Weinberg, in that we primarily aim to create a representation of a fundamental astrophysical concept. 

In particular, we use woven textiles as the medium for our installations, a choice that was inspired by the thread-like appearance of cosmic web filaments (such as those shown in Figure~\ref{fig:viz}). While any work of art must be judged on its own merit, we believe that art, as much as science, seeks to say something true about the nature of existence, and that end is best served by artistic representation that grapples with real data and not only with allegorical concepts. Thus, we try to lay out a stringent process by which we translate simulation data into a textile. Weaving is an especially suitable technique for this project because digitally controlled looms can produce large numbers of thread intersections in an automated fashion. Conversely, the emergence of digital weaving machines in the early 19th century had a critical impact on the development of computers: Jacquard's automated weaving machines were among the first devices to use digital input from punch cards, inspiring the earliest concepts of programmable computers such as Babbage's ``analytical engine.'' By paving the way for digital data input and output, weaving technology helped enable the very computer simulations our work is based on.

What is the motivation behind an arts and science collaboration? From a science perspective, it represents an opportunity to reach an entirely different audience than through conventional outreach such as public talks or planetarium shows. For example, 3D-printed models of astrophysical data have recently been recognized as an important tactile learning tool for the visually impaired \citep[][see also NASA's database of 3D models at \hyperlink{https://nasa3d.arc.nasa.gov/models}{nasa3d.arc.nasa.gov}]{christian_14, grice_15, madura_16}. More fundamentally, artists and scientists share the desire to find and answer new questions and to push current knowledge and methods to their limit. In our quest to connect dark matter with woven textiles, we stretch the limits of current weaving technology and explore new methods to distill a concept like the cosmic web into its very essence. Artists, in turn, express such fundamental notions in novel ways, allowing non-specialists to develop an intuition for otherwise inaccessible concepts.

This paper does not represent a rigorous scientific investigation, but rather a description of our collaboration intended for both artists and scientists. We focus on our creative process and emphasize that the resulting objects and installations do not constitute a final product --- they are intermediate steps along a trajectory of development and experimentation. The paper is laid out as follows. In Section~\ref{sec:sim}, we briefly discuss our $N$-body simulations and analysis methods (this section can be skipped without a loss of continuity). In Section~\ref{sec:3dprint}, we describe our process for translating cosmic web structures into formats suitable for 3D printing and discuss the results. In Section~\ref{sec:fabric}, we transform the cosmic web into a textile sculpture. Throughout the paper, we use some standard astrophysical notation. We denote the density of matter as $\rho$, and the mean density of matter in the universe as $\rhom$. Distances and lengths are generally given in units of $\mpch$, equivalent to $4.66$ million light years or $4.4 \times 10^{19}$ km.

\section{Simulations}
\label{sec:sim}

\begin{figure*}
\centering
\includegraphics[trim = 25mm 25mm 22mm 0mm, clip, scale=0.24]{\figdir/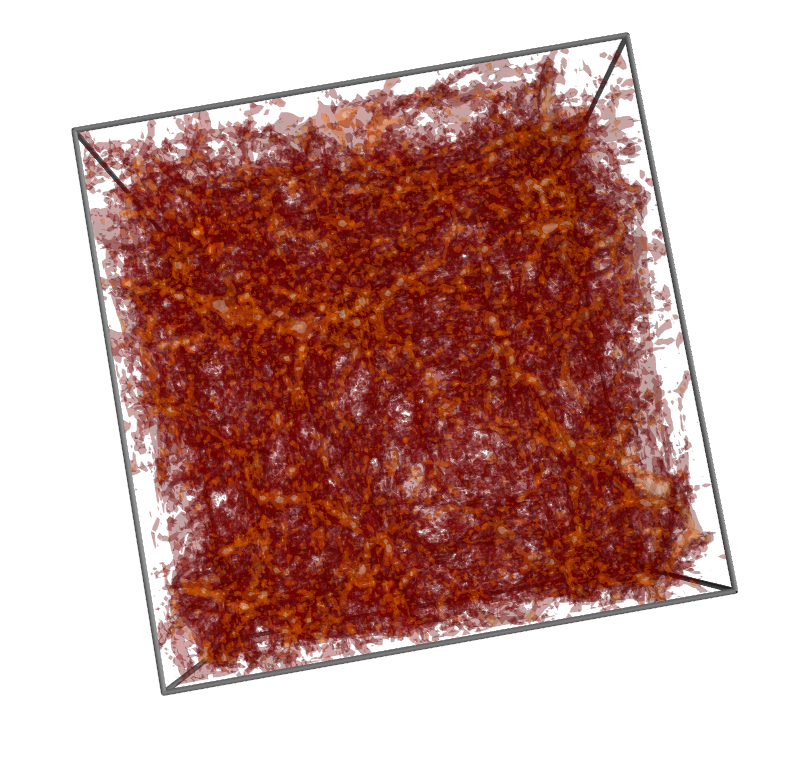}
\includegraphics[trim = 25mm 25mm 22mm 0mm, clip, scale=0.24]{\figdir/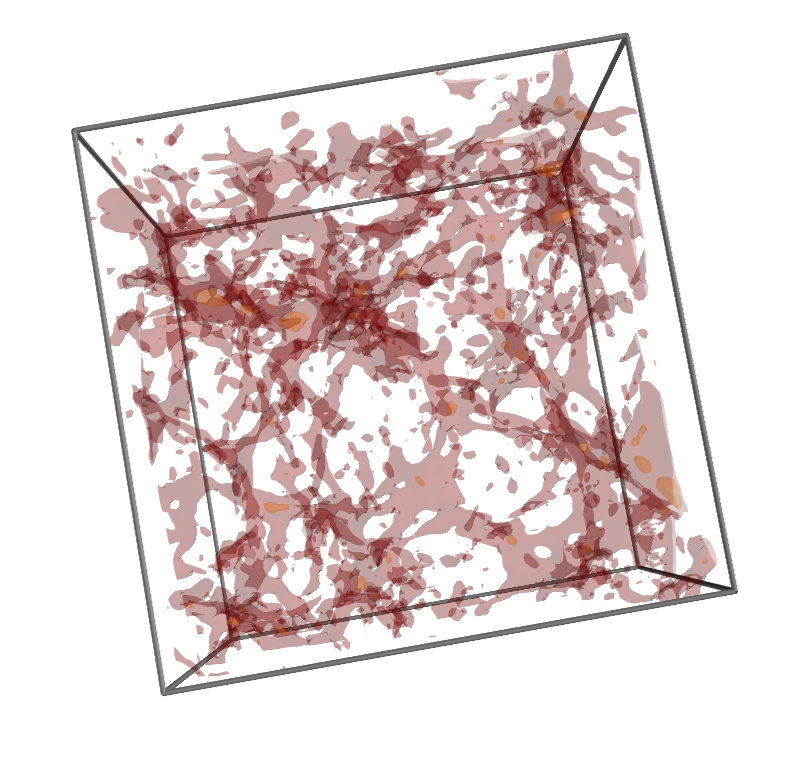}
\includegraphics[trim = 25mm 25mm 22mm 0mm, clip, scale=0.24]{\figdir/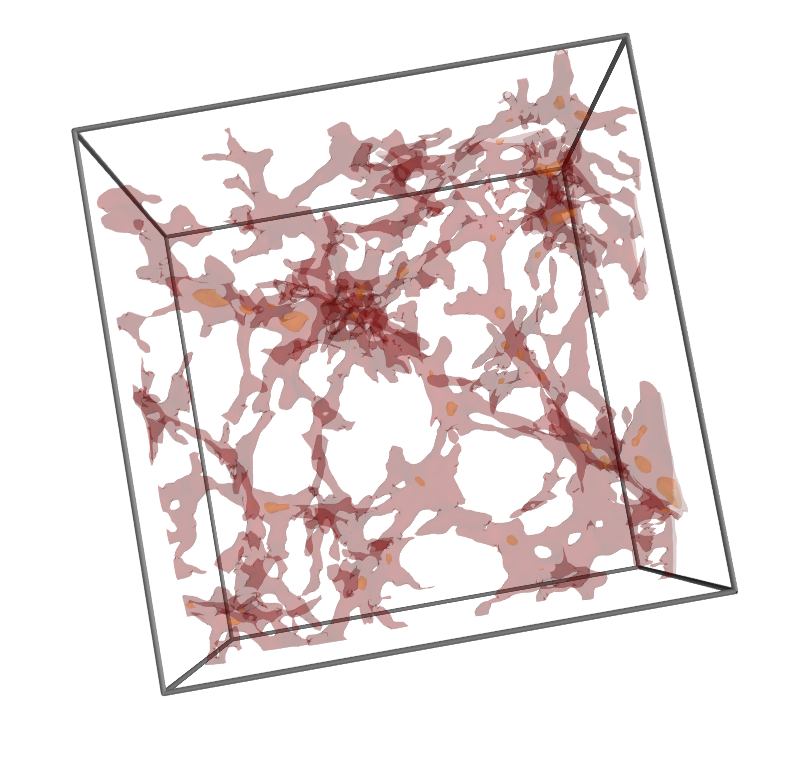}
\includegraphics[trim = 25mm 15mm 22mm 5mm, clip, scale=0.24]{\figdir/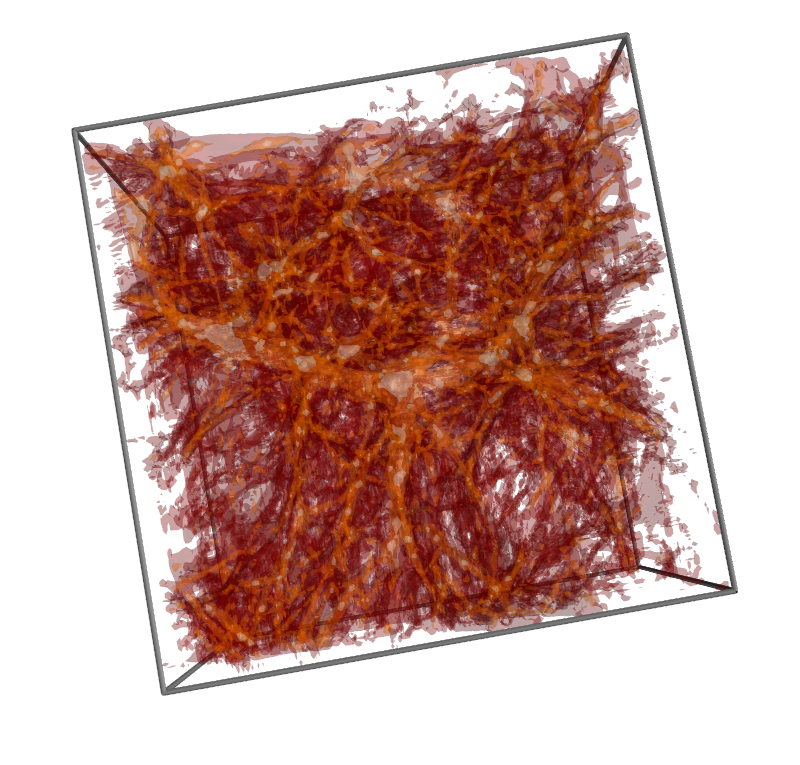}
\includegraphics[trim = 25mm 15mm 22mm 5mm, clip, scale=0.24]{\figdir/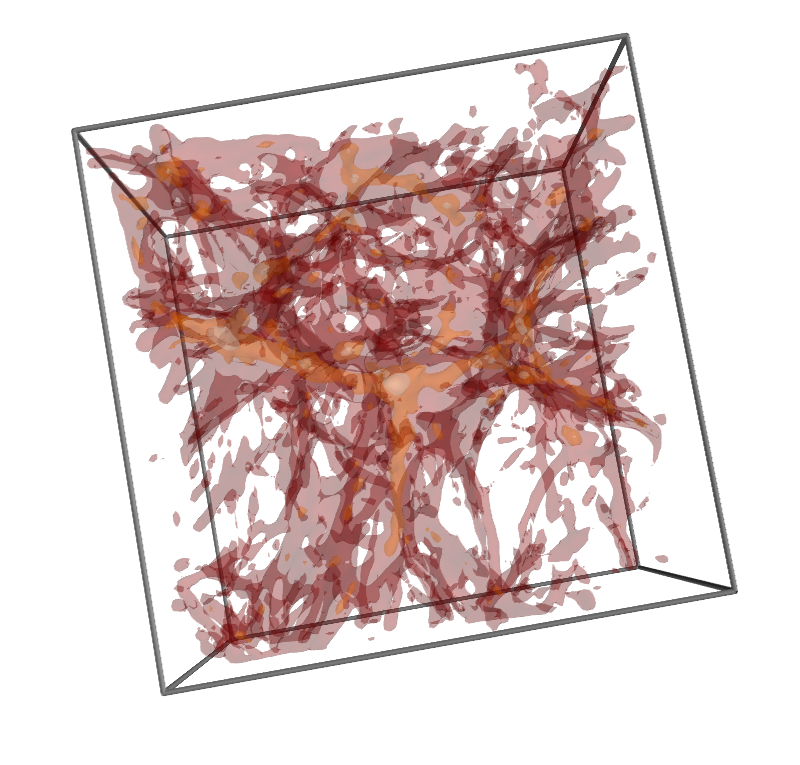}
\includegraphics[trim = 25mm 15mm 22mm 5mm, clip, scale=0.24]{\figdir/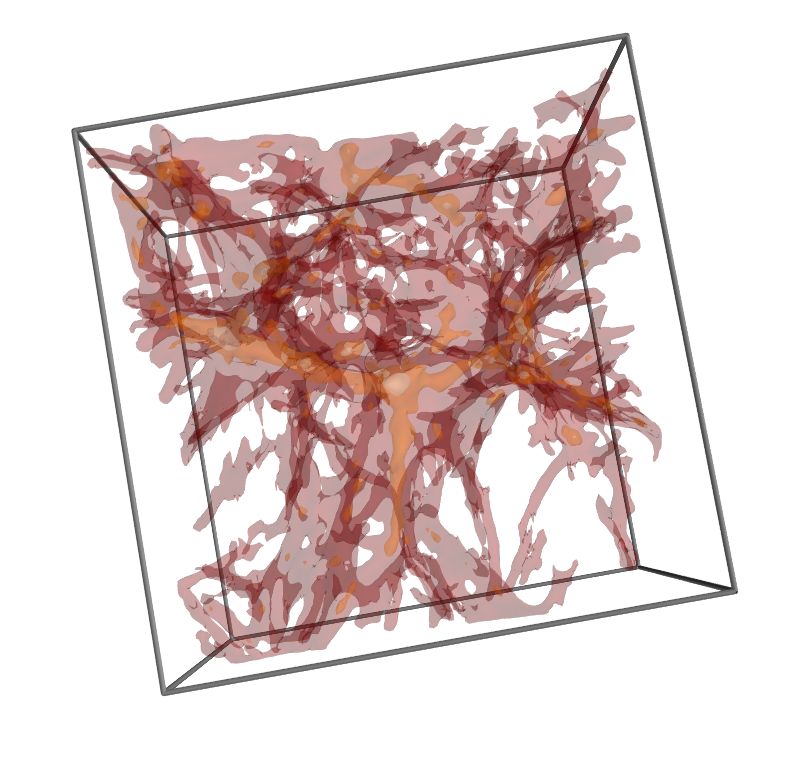}
\caption{Iso-density surfaces of dark matter density in the cosmological volume (top row) and the infall region around a massive halo (bottom row). Three surfaces are super-imposed, namely at densities of $\log_{10} (\rho / \rho_{\rm m}) = 0.1,\ 1$, and $2$ (corresponding to $1.26$, $10$, and $100$ times the mean density of the universe; shown in red, orange, and yellow). The left column shows the density field after minimal smoothing, leading to a complex surface. In the center column, the density field has undergone Gaussian smoothing with a kernel width of one cell. In the right column, unconnected structures have been removed, leaving only the largest coherent structure.}
\label{fig:surfaces}
\end{figure*}

The projects described in this paper are based on an $N$-body simulation of dissipationless dark matter, L0125, which is described in detail in \citet{diemer_14_profiles}. This simulation follows $1024^3$ virtual dark matter particles in a cosmological volume of $125 \mpch$. The L0125 simulation was initialized with the cosmology used in the Bolshoi simulation \citep{klypin_11_bolshoi}, and is consistent with the $WMAP7$ cosmology of \citet{komatsu_11}, a flat \LCDM with $\Omega_{\rm m} = 0.27$. The initial conditions were generated at a starting redshift of $z=49$ using the second-order Lagrangian perturbation theory code \textsc{2LPTic} \citep{crocce_06_2lptic}, and the simulation was run using the publicly available code \textsc{Gadget2} \citep{springel_05_gadget2}. For the textile project described in Section~\ref{sec:fabric}, we use a lower-resolution incarnation of the same simulation with only $256^3$ particles, as higher resolution would have unnecessarily complicated the computations. Once the simulation had been completed, we used the phase--space halo finder \textsc{Rockstar} \citep{behroozi_13_rockstar} to identify halos and subhalos, and the \textsc{Consistent-Trees} code \citep{behroozi_13_trees} to establish subhalo relations and assemble merger trees. 

While halo finding is a common step in simulation analysis, the filaments and walls of dark matter generally receive less attention, largely because the baryonic matter they contain is much harder to detect. Nevertheless, several algorithms for identifying the structural elements of the cosmic web have recently been proposed \citep{aragoncalvo_10, sousbie_11_disperse1, falck_12, shandarin_12, cautun_13, chen_15_cosmicweb, leclercq_15}. Here, we use the public code \textsc{Disperse} which, based on a density field computed from a Delaunay tessellation, applies discrete Morse theory to identify topological features such as walls, filaments, and halos \citep{sousbie_11_disperse1, sousbie_11_disperse2}. 

For the purposes on this paper, we focus on the $z = 0$ output of the L0125 simulation. In particular, we choose two different regions to explore: the entire simulation volume (hereafter referred to as the cosmological volume) as well as a $50 \mpch$ cubic region around a massive cluster halo with mass $\mvir = 4 \times 10^{14} \msunh$ (hereafter called the infall region). Figure~\ref{fig:viz} shows conventional two-dimensional visualizations of these regions. The left panel shows the projection through a slab with a thickness of 10\% of the simulation volume ($12.5 \mpch$), whereas the right panel shows a projection through the entire $50 \mpch$ cube around the infall region. The images were created using the \textsc{GoTetra} code (P. Mansfield et al., in preparation) which computes the density field based on tetrahedra \citep{abel_12, kaehler_12, hahn_13, hahn_15}. In contrast to SPH-like images where particles are smoothed with a kernel function, the tetrahedron method visually emphasizes filaments by assigning density to the regions between particles that were initially close but got separated along a filament.

\section{The Cosmic Web in 3D Prints}
\label{sec:3dprint}

\begin{figure*}
\centering
\includegraphics[trim = 60mm 0mm 40mm 3mm, clip, scale=0.275]{\figdir/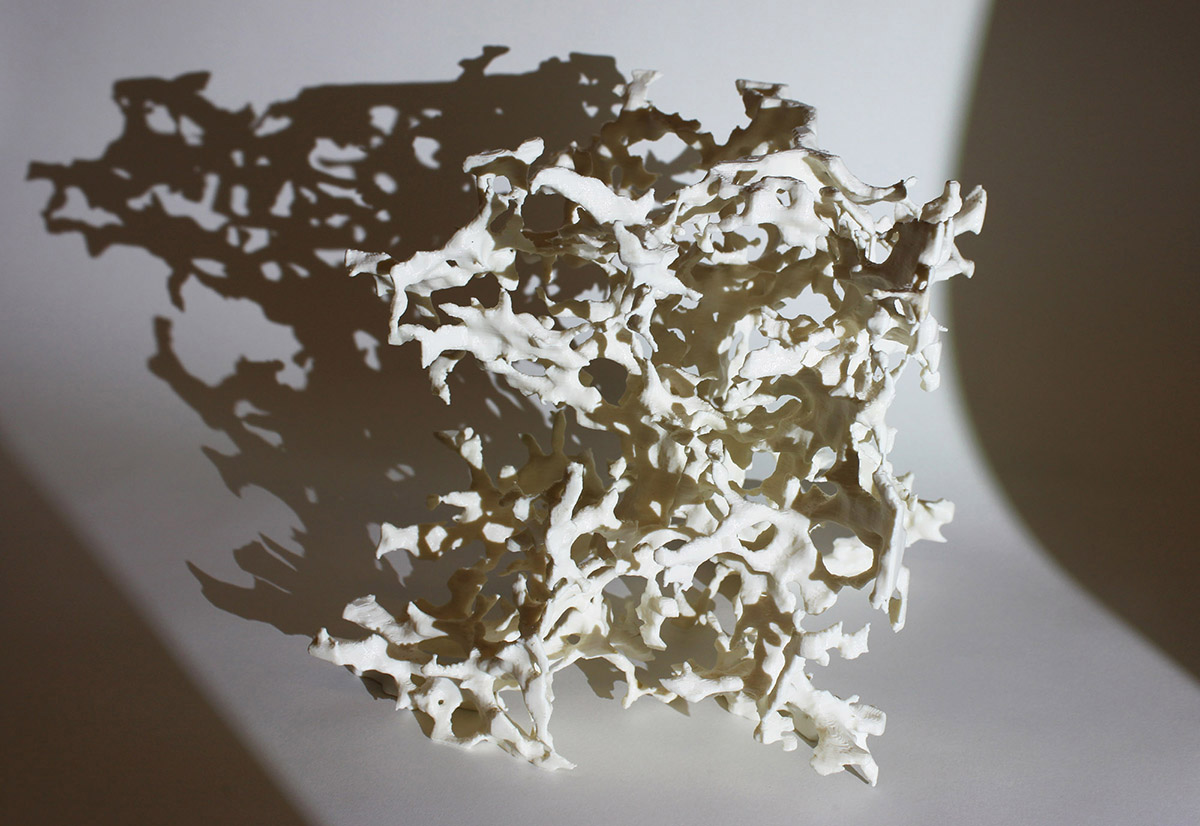}
\includegraphics[trim = 40mm 0mm 60mm 0mm, clip, scale=0.275]{\figdir/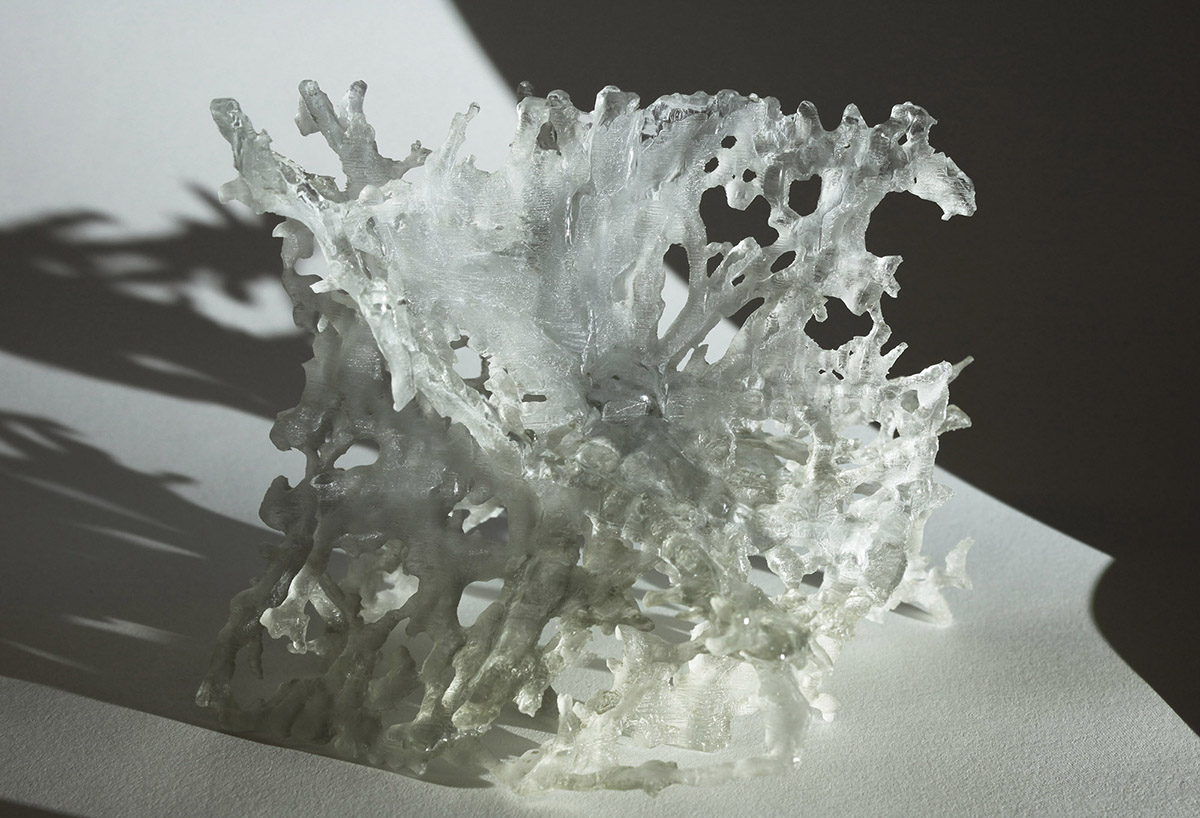}
\caption{3D prints of iso-density surfaces at $\rho = 1.26 \rho_{\rm m}$ for the cosmological volume (left panel) and the infall region around a massive halo (right panel). A prominent wall of infalling dark matter is apparent in the upper part of the latter print, a feature that would be missed in two-dimensional projections of the density field. The prints were produced on different machines and using different plastics, leading to diverse textures and sizes (25 cm and 15 cm, respectively).}
\label{fig:prints}
\end{figure*}

\begin{figure}
\centering
\includegraphics[trim = 50mm 0mm 40mm 0mm, clip, scale=0.34]{\figdir/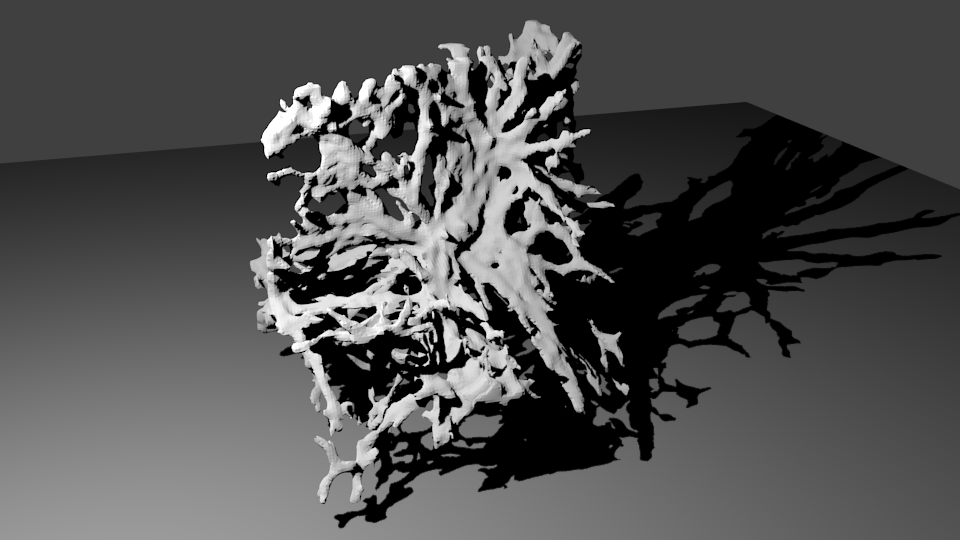}
\caption{Digital rendering of the iso-density surface of the infall region around a massive halo (created using the \textsc{Blender} software).}
\label{fig:render}
\end{figure}

\begin{figure*}
\centering
\includegraphics[trim = 0mm 0mm 0mm 0mm, clip, width=\widththree, height=\widththree]{\figdir/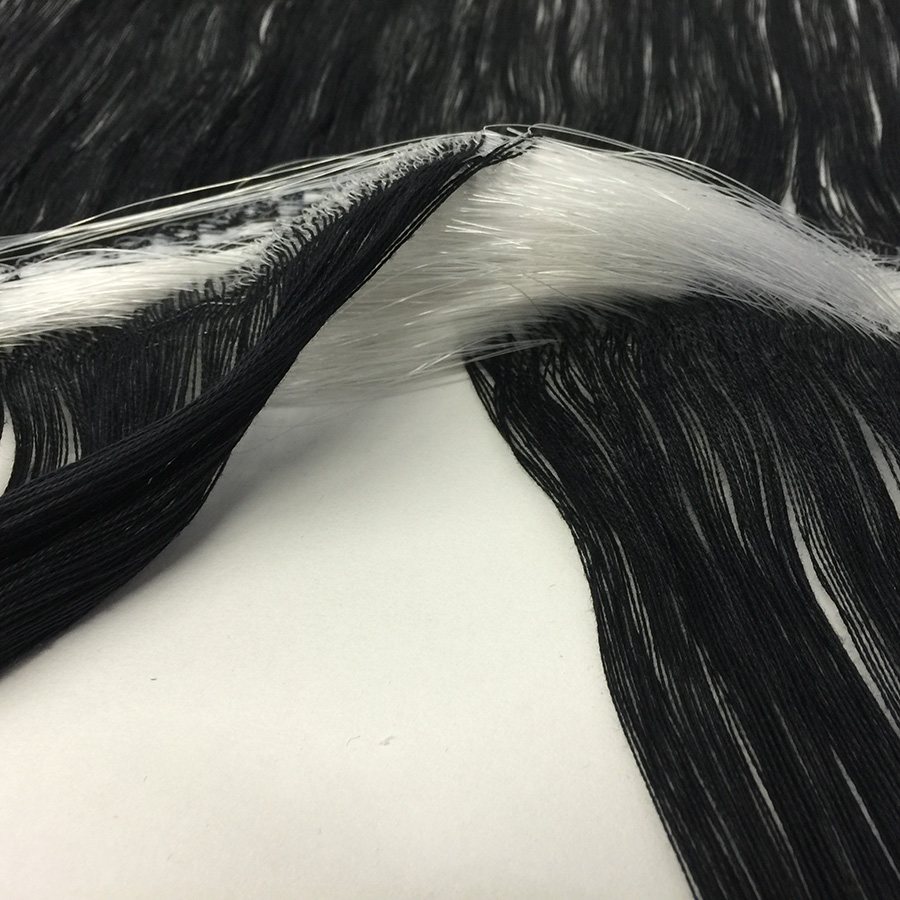}
\includegraphics[trim = 0mm 0mm 0mm 0mm, clip, width=\widththree, height=\widththree]{\figdir/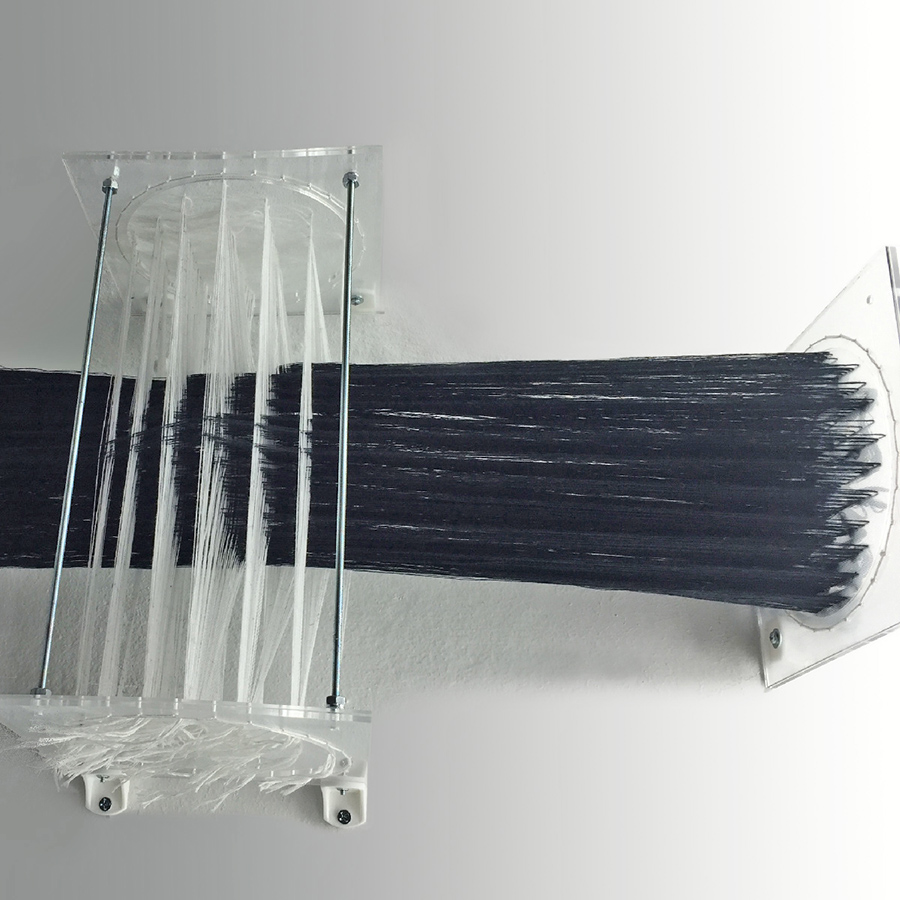}
\includegraphics[trim = 0mm 0mm 0mm 0mm, clip, width=\widththree, height=\widththree]{\figdir/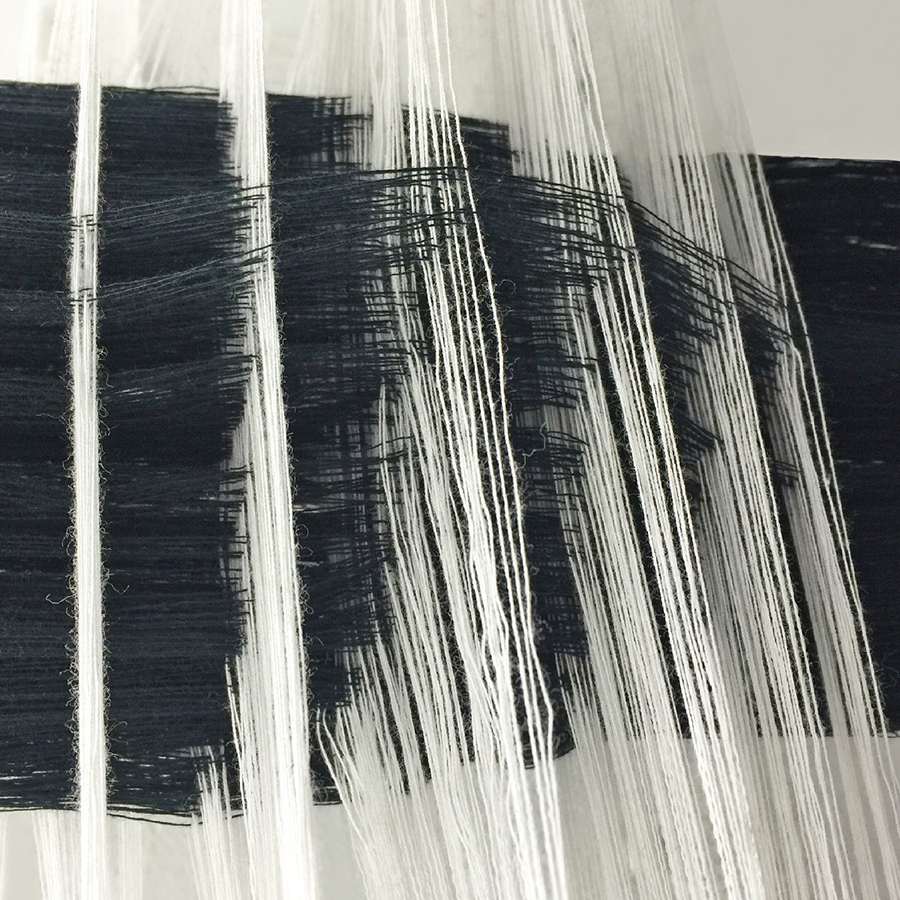}
\caption{Prototypes of woven halos. The left panel shows an early prototype of one quarter of a woven sphere, with black warp and transparent nylon weft threads which form up to 100 layers at the thickest point. The prototype was hand-woven on a TC-1 loom of Digital Norway at the School of the Art Institute of Chicago, a process that took about 12 hr. The right two panels show an example of the type of weave structure used in our textile installation. Here, the white warp and black weft threads are evenly tensioned along 6 warp and 12 weft columns by acrylic panels.}
\label{fig:prototype}
\end{figure*}

\begin{figure}
\centering
\includegraphics[trim = 0mm 0mm 0mm 0mm, clip, scale=0.34]{\figdir/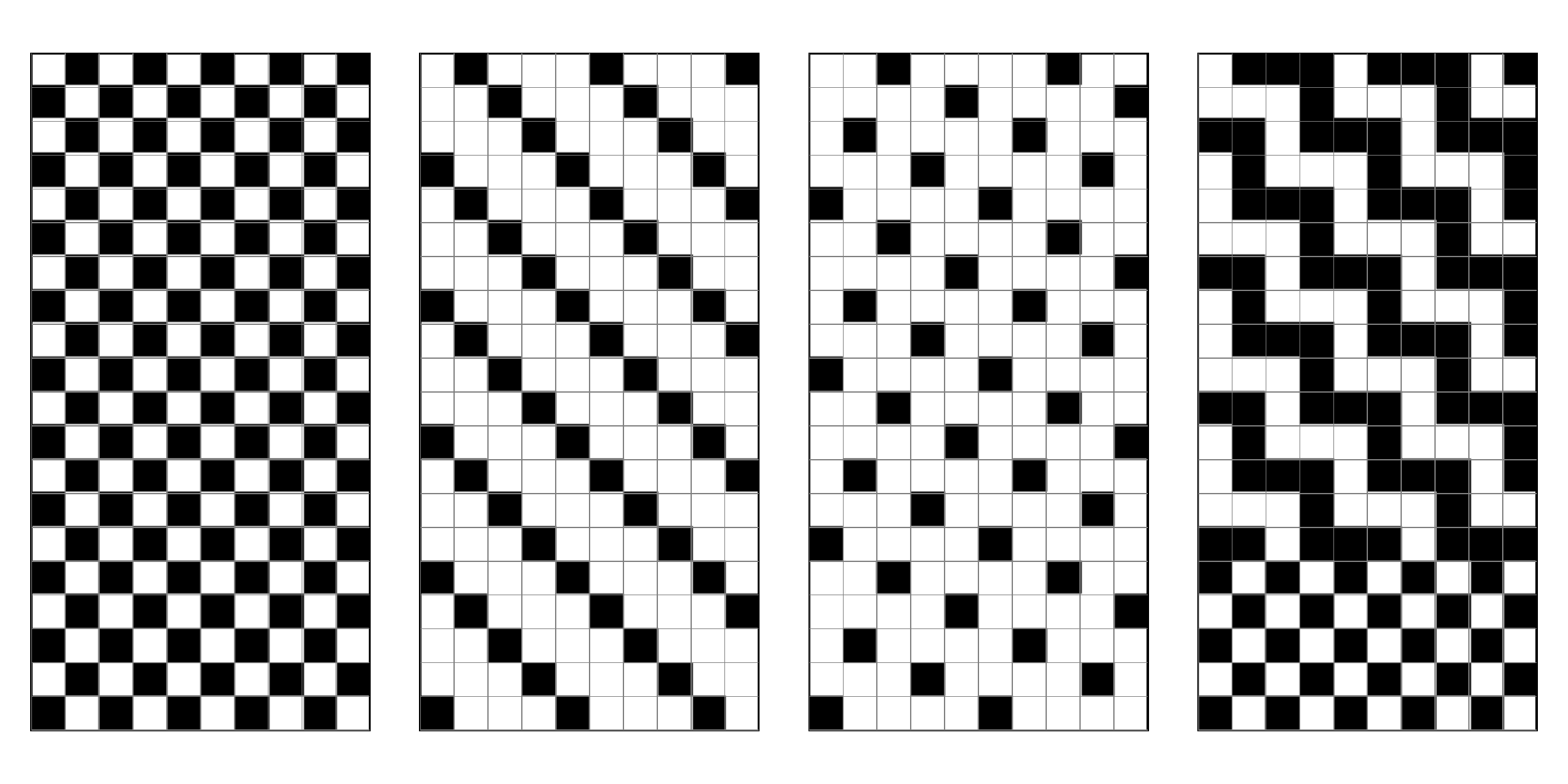}
\caption{Weaving patterns (or ``weave drafts'') for conventional and double-layer cloth. Starting from the bottom, black and white squares represent an up or down position, i.e. whether the threads along the fabric (``warp'') are above or below the threads that run across the width of the fabric (``weft''). The left panel shows a so-called plain weave with an even distribution of warp and weft. The second panel shows a 3:1 twill weave, creating the diagonal pattern that is used for denim Jeans. The third panels shows a satin weave, characterized by a spacing of the binding points by four or more warp or weft threads. All of these basic weave drafts create a single-layer, two-dimensional fabric. The weave draft on the right, however, creates two separate layers. At the bottom, a single layer is woven in plain weave, but then the warp and weft threads separate into two groups that do not intersect, creating two separate layers of plain weave with half the density in each warp and weft.}
\label{fig:drafts}
\end{figure}

Before moving on to more abstract representations of the cosmic web, we wish to gain an intuition for the three-dimensional shapes of typical dark matter structures. Here, the term ``shape'' denotes a surface that can be translated into a tangible object. Images such as those shown in Figure~\ref{fig:viz} do not provide such information: while the human eye picks up on certain patterns in the two-dimensional projection, those impressions are not necessarily accurate. For example, a wall in three dimensions might look like a filament in projection.

We extract density grids with $128^3$ cells using the \textsc{GoTetra} code (though simpler density estimates such as cloud-in-cell would give virtually the same answer). The first grid spans the entire L0125 simulation (Figure~\ref{fig:viz}), giving it a resolution of about $1 \mpch$ per cell. The grid for the infall region has a higher resolution of $0.4 \mpch$. We convert the density grids into logarithmic units of $\log_{10} (\rho / \rho_{\rm m})$ so that $0$ corresponds to the mean density of the universe, $1$ to $10$ times the mean and so on.

Figure~\ref{fig:surfaces} shows three super-imposed iso-density surfaces of the two volumes.\footnote{This visualization was created with the \textsc{mayavi} software which can be downloaded at \hyperlink{http://code.enthought.com/projects/mayavi/}{code.enthought.com/projects/mayavi}.} The naively drawn surfaces (left column) are too complex for the limited resolution of current 3D printers. Thus, we apply a Gaussian smoothing with a standard deviation of one cell to the logarithmic density field. This filter removes much of the small-scale structure (center column of Figure~\ref{fig:surfaces}). While the new surface is much smoother and simpler, we observe a multitude of independent surfaces around local overdensities. Furthermore, the edges of the surfaces touch the boundary of the box, forming open tubes. In order to 3D print an object, it needs to be represented by one ``watertight'' surface, meaning a single, closed surface without any holes. We thus apply two additional corrections. First, we surround the domain by an artificial low-density region so that all surfaces are closed. Second, we identify the largest coherent structure, and discard all smaller structures (right column of Figure~\ref{fig:surfaces}). 

The latter simplification is achieved using the well-known union-find algorithm. We set an arbitrary threshold of $\log_{10} (\rho / \rho_{\rm m}) \geq 0.1$, and check whether each cell satisfies this constraint. If so, we check whether it has a neighbor that also satisfies the constraint, in which case we link the two cells together into a tree. If the other cell has already been assigned a tree root, the new cell is added to the tree, otherwise it begins a new tree. When two trees are linked, all cells belonging to the trees are assigned to one of the tree roots. Eventually, this algorithm produces a set of trees with a certain number of members. We keep the largest tree, and set the density in all other cells to a value below the threshold of the iso-density surface so that they are ignored.

For the 3D prints, we chose surfaces at $\log_{10} (\rho / \rho_{\rm m}) = 0.1$ or $\rho = 1.26 \rho_{\rm m}$. This threshold seems surprisingly low, highlighting how much of the dark matter in the universe is concentrated in spatially small regions. We use \textsc{mayavi} to convert the surfaces into a series of vertices and faces. We then import the resulting wavefront file into the publicly available \textsc{Blender} 3D graphics software\footnote{\textsc{Blender} can be downloaded at \hyperlink{https://www.blender.org/}{blender.org}. \citet{kent_13_blender} and \citet{naiman_16_blender} give further instructions on how to use \textsc{Blender} in an astrophysical context.} and use the 3D printing toolkit plugin to check for any features that might cause trouble during 3D printing. The plugin automatically fixes problematic features such as small holes and very thin bridges. The final cosmological surface has 266,847 vertices and 535,103 faces, the infall region surface has 125,623 vertices and 251,896 faces. We have made these objects (in .stl format) publicly available on our website, \hyperlink{www.fabricoftheuniverse.org}{fabricoftheuniverse.org}.

Figure~\ref{fig:prints} shows photos of the two printed objects. The cosmological volume (left image) was printed on a Dimension SST 1200es printer using opaque plastic, resulting in a maximum side length of 25 cm. The infall region (right image) was printed on an Objet 30 Pro using transparent plastic, resulting in a maximum side length of about 15 cm. The latter print exhibits a prominent infalling wall of dark matter that is not apparent in two-dimensional projections (Figure~\ref{fig:viz}). These features are also visible in a digital rendering created with \textsc{Blender} (Figure~\ref{fig:render}).

\section{The Cosmic Web as a Textile}
\label{sec:fabric}

While we had to make a number of simplifications to create 3D prints, the main goal of that investigation was to obtain as realistic a representation of the simulated density field as possible given the medium. In this section, we describe a more innovative exploration: we investigate the connection between dark matter structures and woven textiles. The objective is not to obtain a realistic depiction of the cosmic web, but rather to push the boundary of both data analysis and weaving technology, and to create a large-scale textile based on simulation data. The central question to this effort is, how can structural information from the cosmic web be translated into a three-dimensional textile? There are multiple, somewhat separate parts to this section. First, we describe the development of a weaving technique that allows us to weave standard, two-dimensional textiles that expand into sphere-like objects resembling dark matter halos. Second, we describe how the density field of dark matter can be transformed into a network of cosmic web elements such as halos and filaments. Finally, we describe how these elements are combined to create a large-scale textile installation.

\subsection{Weaving a Halo}
\label{sec:fabric:weaving}

\begin{figure*}
\centering
\includegraphics[trim = 32mm 10mm 32mm 5mm, clip, scale=0.23]{\figdir/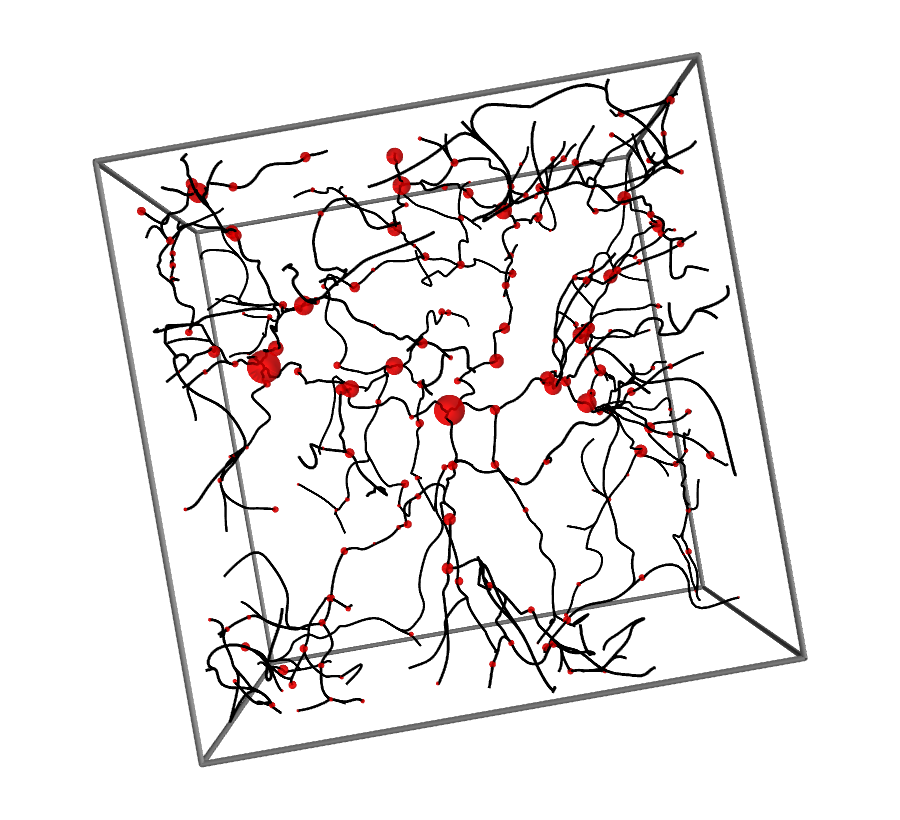}
\includegraphics[trim = 32mm 10mm 32mm 5mm, clip, scale=0.23]{\figdir/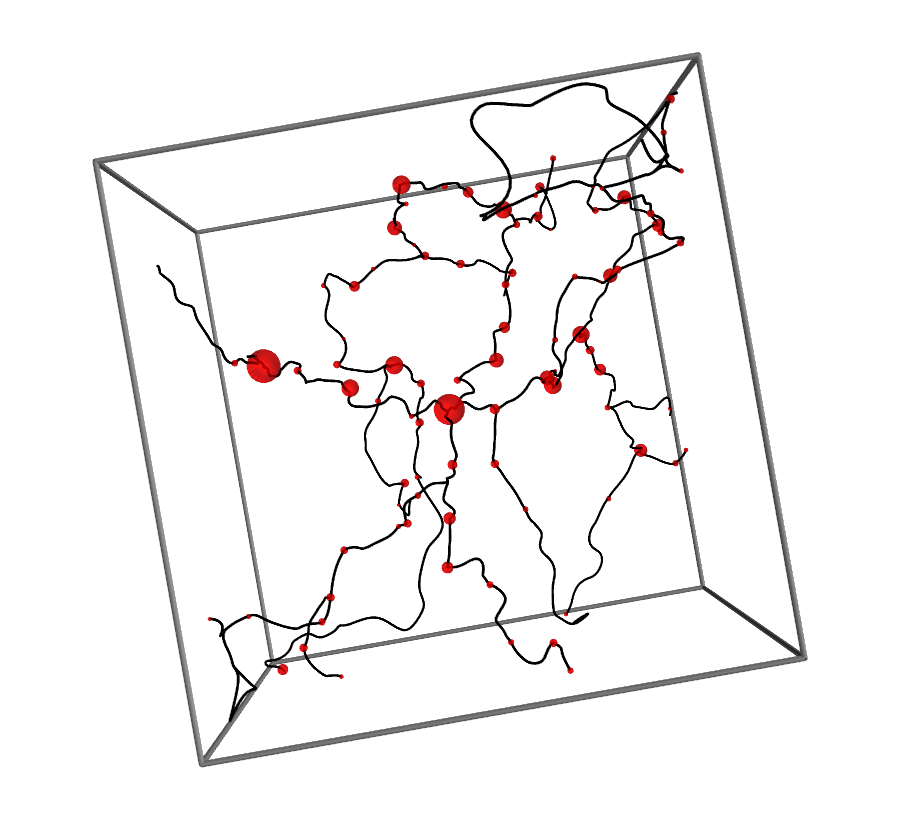}
\includegraphics[trim = 32mm 10mm 32mm 5mm, clip, scale=0.23]{\figdir/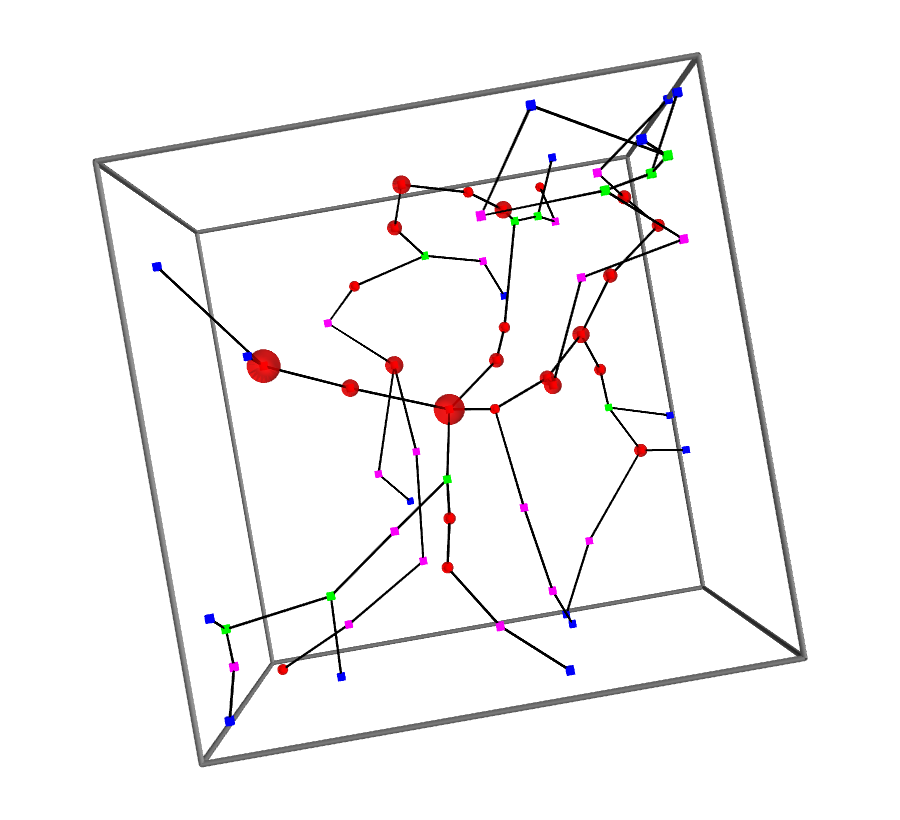}
\caption{Filamentary structure in a $40 \mpch$ region around a massive halo at various stages of simplification. The left panel shows all filaments with a significance above $7 \sigma$ as identified by the \textsc{Disperse} algorithm, with red spheres showing the halos in the halo catalog that lie along those filaments. In the center panel, the significance threshold has been increased to $12 \sigma$. The right panel shows the $12 \sigma$ set simplified into a collection of halos (red), end points (blue), splitting points (green), intermediate points (pink), as well as the straight filament segments between them. The simplification algorithm is described in detail in Section~\ref{sec:fabric:analysis}}.
\label{fig:filaments}
\end{figure*}

A woven fabric is created by interlacing horizontal ``weft'' threads under and over vertical ``warp'' threads (see \citealt{dooley_10_weaving}, \citealt{burnham_81_weaving}, or \citealt{emery_94_weaving} for reviews of basic weaving techniques). Conventionally, this process results in a single-layer, two-dimensional fabric. The weave structure, i.e. the pattern in which the threads are interlaced, is determined by a ``weave draft,'' a sequence of ones and zeros that indicate whether the warp threads are above or below the weft threads. Each change in the up and down positions of adjacent threads creates a binding point. Figure~\ref{fig:drafts} shows a number of typical weave drafts that produce plain, twill, and satin weaves. The right panel shows a weave draft that will lead to a drastically different outcome: two sets of threads are plain-woven at the bottom but then separate from each other, creating two layers that are woven simultaneously but that do not have mutual binding points. This technique can be extended to create large numbers of layers, and eventually a type of three-dimensional fabric. In most applications, multi-layer woven structures are used as a design technique. Here, we push the multi-layering concept to very large numbers of layers that open up into a three-dimensional structure under tension.

In particular, we have developed a weave draft that leads to a spherical shape, reminiscent of a halo. Figure~\ref{fig:prototype} shows two prototypes of this concept. The first (left panel) is a quarter of a sphere woven with black warp and transparent nylon weft threads. At its tallest point, the structure reveals up to 100 woven layers, with fewer and fewer warp threads at the highest layers. The right two panels show an example of the weave structure we eventually used for our installation. The woven sphere is evenly tensioned using acrylic panels. The fundamental weave structure can be scaled to different sizes, we selected four sizes with approximately $140$, $200$, $400$, and $600$ woven layers. The white warp and black weft threads entering the sphere represent the filaments of dark matter that fall into halos in the cosmic web. However, before we can represent the cosmic web as a series of weavings of this kind, we need to transform the dark matter density field into a series of filaments and halos.

\subsection{Breaking the Cosmic Web into Filaments and Halos}
\label{sec:fabric:analysis}

We focus on two structural properties of the cosmic web: the arrangement of filaments, and their relation to halos. We pick a particular region of a particular simulation, translate the three-dimensional density field of dark matter into structural elements such as filaments and halos, and simplify their structure sufficiently for weaving. For this investigation, the resolution of our simulation will make no difference because we have to radically simplify the cosmic web anyway. Thus, we use the smaller $256^3$ particle simulation described in Section~\ref{sec:sim}. We focus on the cubic $40 \mpch$ region around a large halo, similar to the region shown as a 3D print in Figure~\ref{fig:prints}. We use the \textsc{Rockstar} and \textsc{Disperse} codes to extract halos and filaments from our simulation. For our purposes, the \textsc{Disperse} algorithm has one important free parameter: the significance of the extracted filaments, where filaments are removed if they have more than $n \sigma$ chance of appearing in a random density field. Finally, \textsc{Disperse} smooths noisy fluctuations in the filament positions by setting each point to the average of its position and those of its two neighbors. This process was performed $10$ times, ensuring that the filaments are smooth on a scale of roughly $10$ points.

The left panel of Figure~\ref{fig:filaments} shows the set of filaments obtained with a limit of $7\sigma$. Halos are represented as spheres with a size of two virial radii, roughly the physical extent of halos \citep[e.g.,][]{more_15}. The smallest halos are plotted using a fixed size not representative of their actual radius. We only consider halos that overlap with a filament, i.e. where a filament passes within one virial radius of the halo. Furthermore, we exclude the smaller of two halos if their virial radii overlap. 

The filamentary structure shown in the left panel of Figure~\ref{fig:filaments} is somewhat too complicated for our purposes. Thus, we increase the minimum significance of filaments to $12 \sigma$, restricting the data set to the most prominent filaments (center panel of Figure~\ref{fig:filaments}). Due to the criteria described above, this reduces the number of halos as well. However, the dataset is still not suitable for translation into a woven fabric. First, we cannot create arbitrary shapes in three-dimensional space, but are limited to a collection of intersection points (halos) and straight bundles of thread (filaments). Furthermore, the number of points that can realistically be fixed in three-dimensional space in an installation is limited. Thus, we translate the filament data set into a set of four types of anchor points (right panel of Figure~\ref{fig:filaments}): halos (red), end points (blue), branching points (green), and intermediate points in long filaments (magenta). We create this set of points by first placing the halo anchor points, namely all halos above a certain mass through which some filament passes. Second, we add the end points of filaments, but whenever a new anchor is within a certain distance of an existing anchor, (e.g., a halo), we merge the two for simplicity. Third, the \textsc{Disperse} algorithm outputs filaments that may share large fractions of their length with one another. We identify which filaments overlap, and insert an anchor point wherever they stop overlapping (again preferring already existing anchors if possible). Finally, some long stretches of filament may deviate far from the original filaments if they are not supported by an anchor point. Thus, we place intermediate anchors wherever a filament exceeds a certain length without encountering an anchor point.

We have now simplified the structure sufficiently to attempt translating it into a textile. However, the weaving is performed on two-dimensional looms, meaning we need to translate the three-dimensional structure shown in the right panel of Figure~\ref{fig:filaments} into a two-dimensional diagram (Figure~\ref{fig:tree}). In this step, we have introduced a additional simplifications. First, we have broken a few closed loops of filament as they would be hard to realize in a weaving (the loops are apparent in the right panel of Figure~\ref{fig:filaments}). Second, we have replaced all split points with a small halo, since halos will represent our way of binding filaments together. Physically speaking, this is not unrealistic: a halo is likely to be located at intersections of filaments, but might have been omitted because of our mass cut. Third, we have binned the size of the halos into only four sizes in order to reduce the number of different woven objects that need to be produced. The sizes are indicated with different shades of red in Figure~\ref{fig:tree}. At this point, our algorithm has become relatively complicated but performs the simplification without human interference.

\subsection{Assembling the Textile Installation}
\label{sec:fabric:installation}

\begin{figure}
\centering
\includegraphics[trim = 75mm 27mm 49mm 80mm, clip, scale=0.60]{\figdir/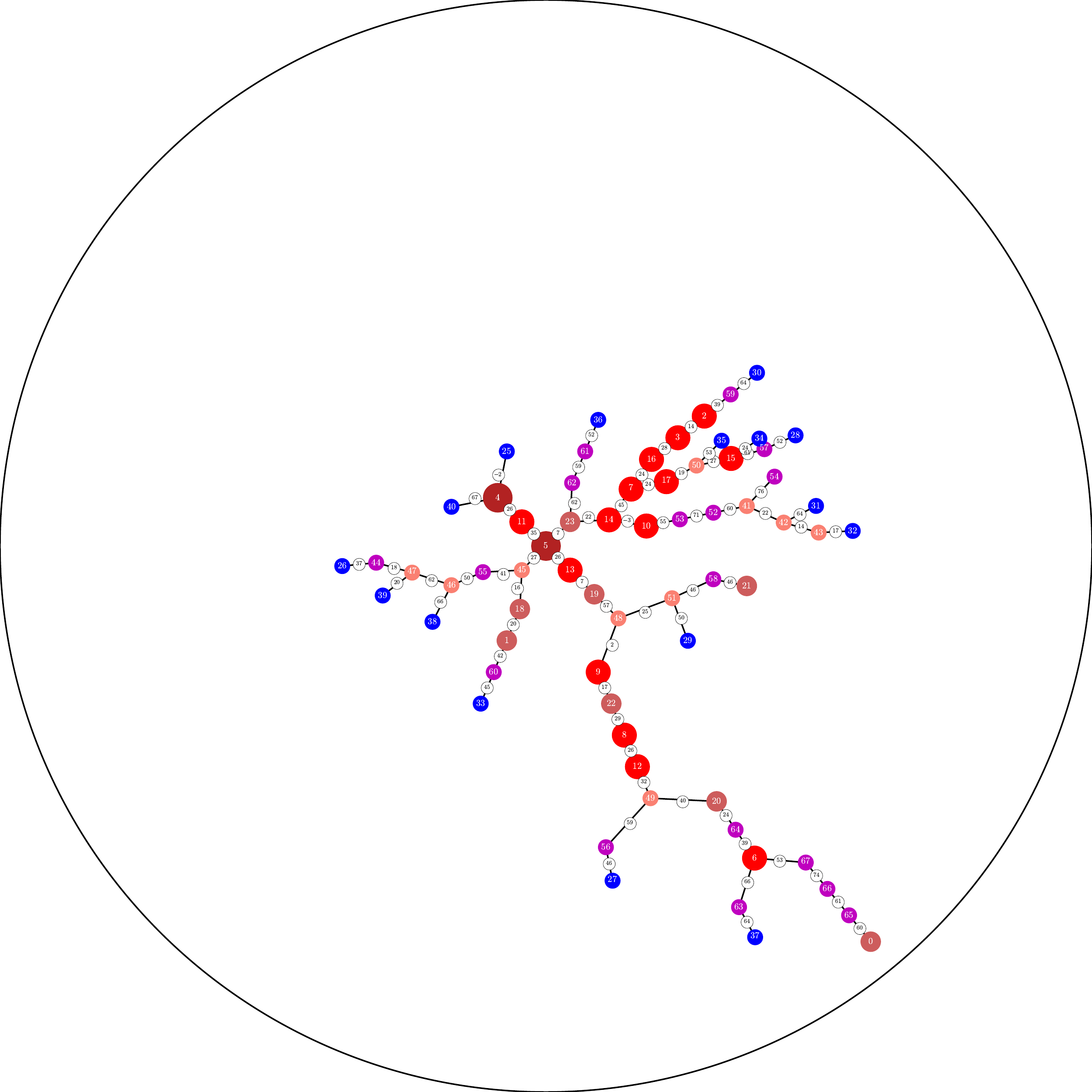}
\caption{Tree diagram representation of the simplified filamentary structure. Red points represent halos (with different shades indicating different sizes), blue points represent end points, and purple points intermediate filament points. The ID numbers for each anchor are arbitrary identifiers used when assigning the woven structures a location in three-dimensional space. The numbers in the white circles give the lengths of filament required between each set of anchor points (in cm). This diagram represents the final automated step in our analysis pipeline before the structure is translated into a weave pattern.}
\label{fig:tree}
\end{figure}

At this point, we need to make some aesthetic choices regarding the physical textile we aim to produce. First, we fix the size of the structure to a cube of 300 cm side length, corresponding to a scale of $7 \times 10^{-23}$. For comparison, the disk of the Milky Way galaxy would be about 1 mm wide at this scale, and the solar system about the size of an atom. Figure~\ref{fig:tree} shows the lengths of filament corresponding to this total size in white circles. We note that we could have chosen any shape of volume, but we prefer the cubic volume which is reminiscent of the cubic domain typically chosen for cosmological simulations.

\begin{figure}
\centering
\includegraphics[trim = 0mm 0mm 250mm 270mm, clip, scale=0.07]{\figdir/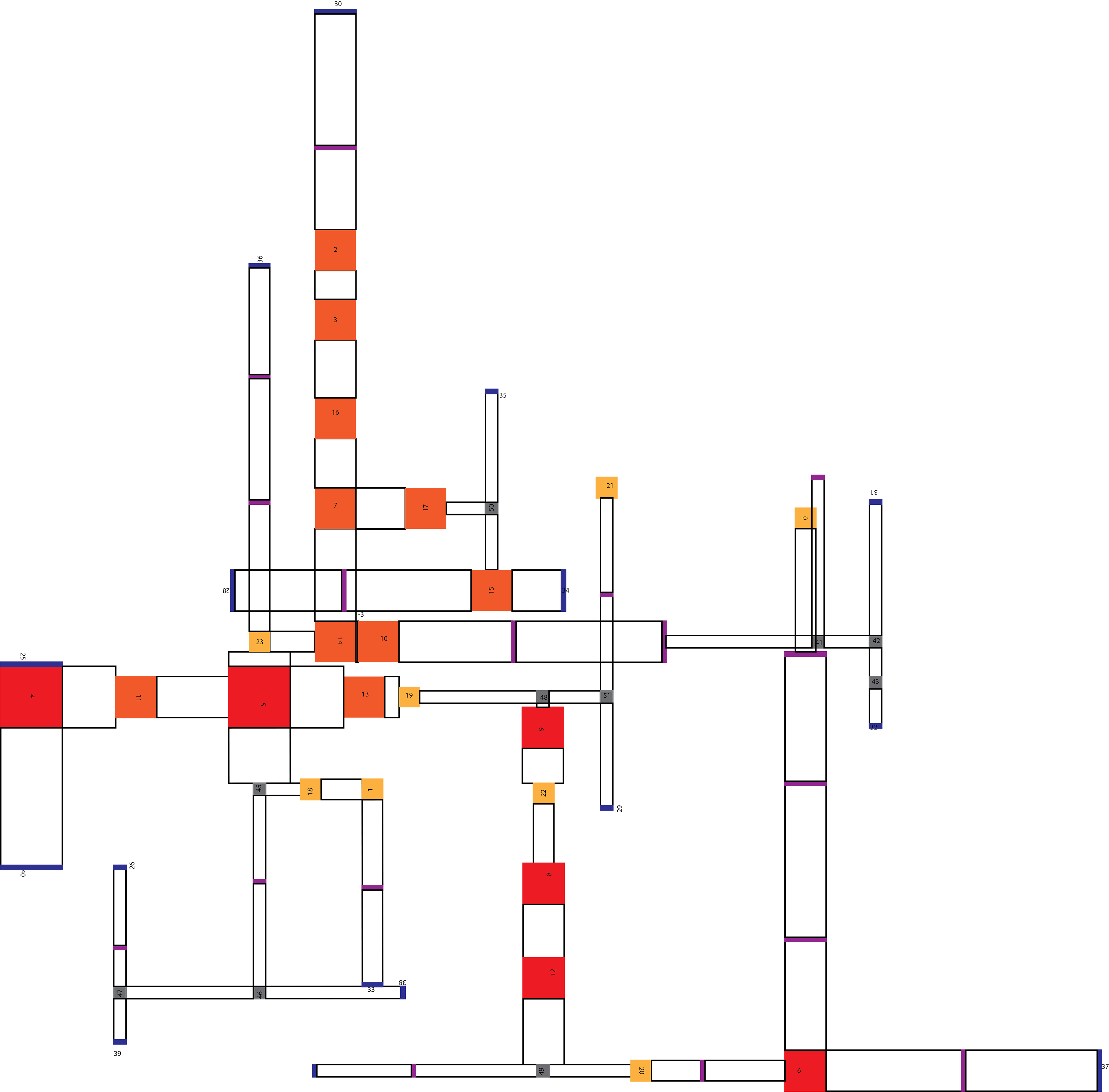}
\caption{Layout of the textile as a large weave draft, a direct translation of the tree diagram in Figure~\ref{fig:tree}. Halos of different sizes are marked in red, orange, yellow, and green. Filaments appear as white stripes between them and represent regions without any binding points (or ``floats''). The blue end points of filaments represent places where filaments must be attached (e.g., to a wall). The layout had to be separated into multiple sections due to the limited width of the weaving loom (about 150cm).}
\label{fig:layout}
\end{figure}

Given the tree diagram in Figure~\ref{fig:tree}, we translate the filament structure into a two-dimensional weave draft. Due to the practical constraints of the weaving loom (e.g. a limited width and a different density of points in the warp and weft direction), this step was performed manually. Figure~\ref{fig:layout} shows part of the resulting schematic. From this schematic, we generate the weave draft that is fed to the weaving machine, i.e. we replace the colored rectangles that represent halos with the halo weave drafts discussed in Section~\ref{sec:fabric:weaving}.

The textile was woven on an industrial Dornier Jacquard weaving machine at the TextielLab in Tilburg, Netherlands. The resulting piece of fabric was 12 m long and took about 8 hr to weave. The largest halos contain up to 600 vertically stacked layers at their thickest point. We chose to use white thread for both the warp and weft, though reflective thread was substituted for one in $50$ weft threads to create a glow when the fabric is illuminated. The individual filaments were mounted in a circular distribution on laser-cut acrylic disks in order to put tension on the halos. The disks, in turn, were attached to walls and about 30 vertical aluminum stands, mounted on steel feet. The positions and heights of the stands were accurately measured in order to reproduce the three-dimensional structure of the filaments and halos in the simulation.

\begin{figure*}
\centering
\includegraphics[trim = 0mm 0mm 0mm 0mm, clip, width=16.07cm]{\figdir/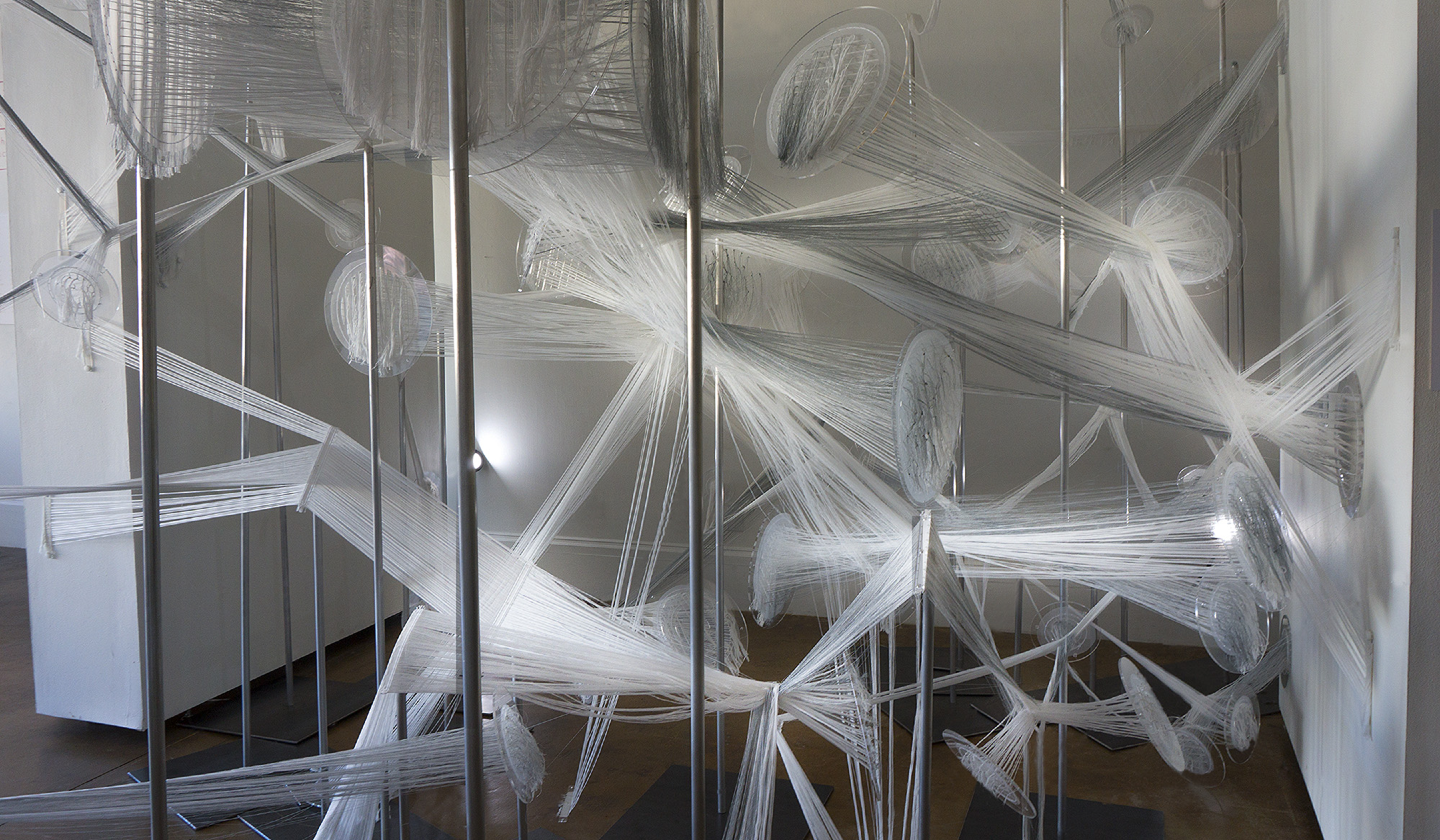}
\includegraphics[trim = 0mm 0mm 0mm 0mm, clip, height=8cm]{\figdir/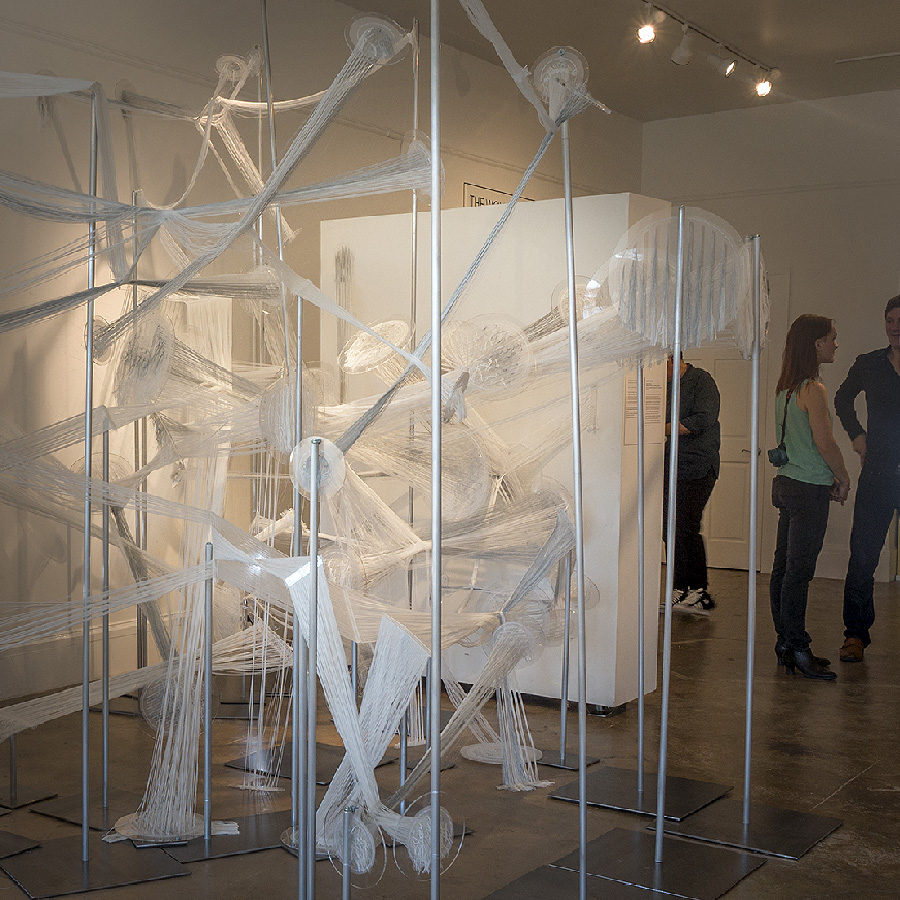}
\includegraphics[trim = 0mm 0mm 0mm 0mm, clip, height=8cm]{\figdir/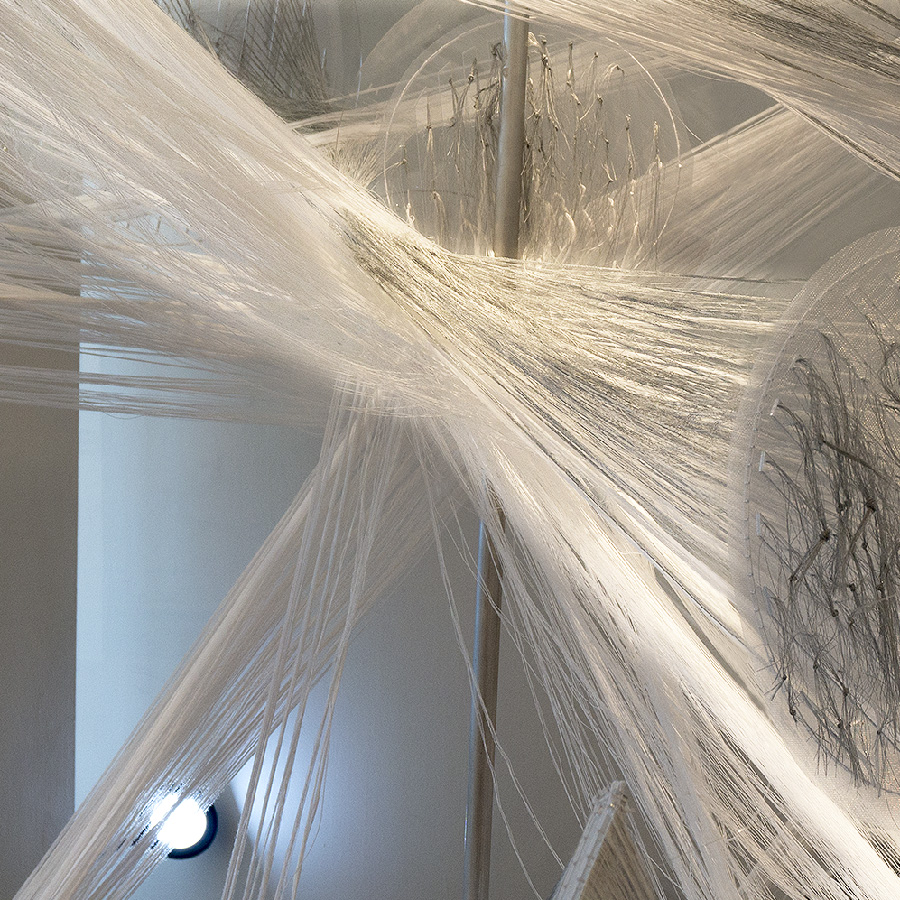}
\caption{First large-scale textile installation at the Non-Fiction Gallery in Savannah, GA. Halos are represented as the woven intersection of filaments, parallel bundles of thread. Walls and vertical stands hold the filament end points and turning points in the appropriate three-dimensional positions. The two images at the bottom right zoom in on particular woven halos. As the warp and weft thread filaments are pulled in the directions dictated by the underlying three-dimensional structure, the woven halos take on drastically different shapes. Photos by David Martinez-Moreno.}
\label{fig:installation}
\end{figure*}

The textile was exhibited at the Non-Fiction Gallery in Savannah, GA in October 2016. It was part of an exhibition entitled ``The Woven Cosmos'' in the context of the 15th Biennial Symposium of the Textile Society of America. Figure~\ref{fig:installation} shows images of the installation.

\subsection{Future Plans}
\label{sec:fabric:future}

In the process of developing and building the Savannah textile installation, we discovered a number of challenges, not all of which could be addressed in this first large-scale weaving. Most importantly, all halos have four filaments attached to them (two sets of each warp and weft threads), even if the simulated data set prescribes fewer filaments. In order to keep tension on the woven halos, both the intended and spurious filaments had to be mounted, creating a visual distraction from the shapes shown in Figure~\ref{fig:filaments}. Similarly, the aluminum stands and acrylic disks were necessary for technical reasons, but visually distract from the textile itself.

We hope to improve on these issues in future installations. In particular, we aim to create a textile that is suspended from walls and ceiling, or supported with less distracting design elements. Moreover, we will experiment with multi-color weavings to create a visual contrast between intended and spurious filaments. More information, visual materials such as videos and images, as well as current updates on our projects can be found on our website, \hyperlink{www.fabricoftheuniverse.org}{fabricoftheuniverse.org}.


\vspace{0.5cm}

\section*{Acknowledgments}
\label{sec:ack}

This project could not have been realized without the help of numerous individuals and organizations. First, we are deeply grateful to those who helped us create, namely: the TextielLab/TextielMuseum in  Tilburg, Netherlands, with their invaluable help in designing our weavings and producing our textile, in particular Judith Peskens, Hebe Verstappen, and Errol van de Werdt; our indefatigable studio assistant Katerina Williams who mounted the textile; the team that designed and built the Savannah exhibition, namely Helen Yuanyuan Cao, David Martinez-Moreno, Heather MacRae, Heather McKenzie, Julie Ann Miller, Elizabeth Pope, and Susanna Kim Vitayaudom; the Advanced Output Center at SAIC who produced our 3D prints; Harvard MCB graphics who produced large-scale prints for our exhibition; and the Research Computing Center (RCC) at the University of Chicago who provided the computational resources used in this project. Finally, we thank the individuals who provided the software used in this project, in particular the codes \textsc{Disperse} (Thierry Sousbie), \textsc{Gadget2} (Volker Springel), \textsc{GoTetra} (Philip Mansfield), and \textsc{Rockstar} (Peter Behroozi).

The financial support from a number of organizations made this project possible, namely the Arts, Science, and Culture Initiative at the University of Chicago whose Graduate Collaboration Grant initiated this project. We thank Julie Marie Lemon and Marissa Benedict for their support and encouragement in the early stages of this project. We are also grateful to the Earl and Brenda Shapiro Center for Research and Collaboration at the School of the Art Institute for an Early Concept Research Grant for Exploratory Research, with special thanks to Rebecca Duclos and Jaclyn Jacunski. 

We thank our advisors Michiko Itatani, Kathleen Kiefer, Andrey Kravtsov, Susan Snodgrass, Lynn Tomaszewski, and Francis Whitehead, as well as our home institutions, namely the Department of Fiber and Material Studies, the Textile Technology Research Group, the Dean's office, and the Department of Textiles at the School of the Art Institute of Chicago, as well as the Art Institute of Chicago, the Department of Astronomy and Astrophysics at the University of Chicago, and the Institute for Theory and Computation at the Harvard-Smithsonian Center for Astrophysics.

Finally, we are grateful to Kathleen Kiefer, Jill Naiman, Lynn Tomaszewski, Mika Turim-Nygren, and the anonymous referee for comments on a draft of this paper. 


\bibliographystyle{aasjournal}
\bibliography{../../../_LatexInclude/sf.bib}

\end{document}